\documentstyle[11pt,aaspp4]{article}

\lefthead{Smith et al.}
\righthead{HD19994} 

\begin{document}

\title{The Abundance Distribution in the Extrasolar-Planet Host Star
HD19994}

\author{Verne V. Smith}
\affil{Department of Physics, University of Texas at El Paso, El Paso, 
TX 79968  USA, and McDonald Observatory, University of Texas at Austin,
Austin, TX 78712 USA; verne@barium.physics.utep.edu}

\author{Katia Cunha}
\affil{Observat\'orio Nacional, Rua General Jos\'e Cristino 77, 
20921-400 S\~ao Crist\'ov\~ao, RJ, Brazil; katia@on.br}

\author{Daniela Lazzaro}
\affil{Observat\'orio Nacional, Rua General Jos\'e Cristino 77, 
20921-400 S\~ao Crist\'ov\~ao, RJ, Brazil; lazzaro@on.br} 

\begin{abstract}

Abundances of 22 elements have been determined from a high-resolution,
high signal-to-noise spectrum of HD19994, a star recently announced
as harboring an extrasolar planet.  A detailed spectroscopic analysis
of this star finds it to have a mass of 1.2$\pm$0.1M$_{\odot}$.
HD19994 is found to be slightly enriched in ``metals'' relative to
the Sun ([Fe/H]=+0.09$\pm$0.05 and an average of all elements
of [m/H]=+0.13), as are most stars known with extrasolar planets. In an
investigative search for possible signatures of accretion of metal-rich
gas onto the parent stars in such systems (using HD19994 and published
abundances for other stars), it is found that a small subset of stars
with planets exhibit a trend of increasing [X/H] with increasing
condensation temperature for a given element X.  This trend may point
to the accretion of chemically fractionated solid material into the outer
(thin) convection zones of these solar-type stars. 
It is also found that
this small group of stars exhibiting an accretion signature all have
large planets orbiting much closer than is found, in general, for stars
with planets not showing this peculiar abundance trend, suggesting 
a physical link between accretion and orbital separation.  In addition,
the stars showing evidence of fractionated accretion are, on average,
of larger mass (1.2M$_{\odot}$) than stars not showing measurable
evidence of this accretion (1.0M$_{\odot}$).  

\end{abstract}

\keywords{planetary systems --- stars: abundances}
\newpage
\section{Introduction}

The last five years have been witness to a continuing series of detections 
of extrasolar planets around solar-type stars 
using the Doppler method.  The number of systems discovered continues to
grow, with the current number of solar-type stars known to have planets
approaching 50; the first detections were by Mayor \& Queloz (1995), Butler
\& Marcy (1996), and Marcy \& Butler (1996), with recent systems being
announced by Queloz et al. (2000) and Butler et al. (2000).  The 
current sample of
solar-type stars known to harbor planetary systems is biased, of course,
by the Doppler method being most sensitive to large planets, as well as
large planets that closely orbit their parent stars (neglecting variations 
in stellar mass, which do not differ greatly over the limited range in
spectral type from K to F dwarfs).  Thus, the systems detected to date are 
probably not a representative sample of all planetary systems around 
solar-type stars.  Nonetheless, much can potentially be learned about the
formation and evolution of planetary systems by focussing on the detailed
properties of the known stars with (large) planets. 

One striking property of the stars with planets that have been
analyzed spectroscopically to date is their metallicity distribution,
where the iron abundance is used typically as a fiducial element 
with which to define an
overall stellar metallicity.  As noted first by Gonzalez (1997, 1998), and
largely confirmed by continuing work (Gonzalez \& Vanture 1998; Gonzalez,
Wallerstein, \& Saar 1999; Gonzalez \& Laws 2000), the Fe abundances
in stars with planets tend to be larger than typical values found for
field F, G, or K stars not known to have planets.  This trend is
confirmed most recently in work by Gonzalez et al. (2001) and Santos,
Israelian, \& Mayor (2001).  
The differing Fe abundances found between a sample of stars with
known (large) planets and a sample of solar-type dwarfs (presumably)
more representative of the Galactic disk, suggests that the presence of
large planets and Fe abundance, or more generally metallicity, are related.
Two possible connections have been discussed: either the
presence of planetary systems can influence the metallicity of parent
stars or, on the other hand, planets may form preferentially
in more metal-rich environments.  Discriminating between these two 
possibilities will help in understanding how systems with large planets
form. 

In the first case, it is argued that a parent star in a planetary
system might accrete
chemically fractionated material, depleted in H and He, during 
early evolution in the system when the star is surrounded by a disk.
The mass in the outer convection zone of solar-type stars is relatively 
small throughout the main-sequence lifetime and material deposited in the
envelope 
can remain there.  Accretion of a substantial amount of H- and He-depleted
material, relative to the mass in the stellar convection zone, could
enhance the metallicity of the outer mass zone of the parent star.
The scenario of a young star accreting substantial amounts of material from
a protoplanetary disk became a distinct possibility with the discovery of 
large planets with small orbits, as typified by 51 Peg and $\tau$ Boo.
It is now suggested (e.g. Lin et al. 1996; Ward 1997; Trilling
et al. 1998) that large planets observed close to their parent
stars actually formed at larger distances, but migrated inwards
due to combinations of drag and/or tidal forces.   During inward migration,
material from the disk, depleted in H and He, could be accreted onto
the star. 

Another possibility is that an elevated metal abundance in the natal cloud
out of which a star forms might enhance the probability of forming a large
planet.  According to the standard model (Mizuno 1980; Bodenheimer \&
Pollack 1986; Pollack et al. 1996; Burrows et al. 1997), the formation
of Jovian-mass, or larger, planets is initiated by the condensation of
``ices'', such as H$_{2}$O or CO$_{2}$, thus the primordial abundances of
such elements as C or O in the molecular cloud may play a role in
large-planet formation.   

Quite probably, the processes involved in planet formation are
complex, with a number of variables dependent on both primordial
abundances and accretion.  Detailed chemical abundance analyses 
of stars known to possess planetary-mass companions will provide
useful data in the investigation of how the systems with large
planets have formed.  In this study, an abundance analysis of 22 elements
has been carried out for HD19994 (94 Cet, HR962, HIP14954), which was
recently announced by Queloz et al. (2001) to have a planet.  This planet 
has a mass of M$_{\rm P}$sin$\iota$= 2.0M$_{\rm Jupiter}$, orbits
with a semi-major axis of 1.3AU, and has a period of 454 days.  The
chemical abundances in HD19994 are compared to abundances in other
planet-harboring stars studied by Gonzalez et al. (2001) and Santos
et al. (2001). 

\bigskip
\section{Observations}

The spectrum analyzed here was taken at the University of Texas' McDonald
Observatory with the Sandiford cross-dispersed echelle spectrometer, 
attached to the cassegrain focus of the 2.1m Struve reflector.
This spectrometer provides a two-pixel resolving power of 
$\lambda$/$\Delta$$\lambda$=R=60,000
on a 400 x 1200 pixel CCD detector and the spectral coverage for these data 
were from 6050--7900\AA.  The star was observed in February 2000.  The
actual measured resolution achieved in this spectrum of HD19994 was 2.2 pixels
(full-width half-maximum), or R= 54,500. 

Data reduction utilized the IRAF software from NOAO.  Bias CCD frames
were subtracted from the raw program CCD images (star, internal
quartz flat-fields, and Th--Ar hollow cathode lamps).  The two-dimensional
locations of the spectral orders were defined, interorder light was
then identified and polynomial fits to this light were
made in both the dispersion and cross-dispersion directions, with
the resultant interorder light subtracted from the image in question.
The bias subtracted and interorder corrected frames were divided
by the flat-field images and the defined spectral orders were 
summed and extracted to obtain a set of one-dimensional spectra.  Wavelength
calibrations were set from the Th--Ar spectra, with typical residuals
from a wavelength fit of 5--8 m\AA.  The final signal-to-noise ratio, measured
near 6500\AA, was S/N$\sim$ 400. 

\bigskip      
\section{Analysis}

The techniques employed to derive abundances center on using the spectrum
to derive the fundamental stellar parameters effective temperature 
(T$_{\rm eff}$), surface gravity (parameterized as log g), microturbulent
velocity ($\xi$), and metallicity (characterized by [Fe/H]). The model
atmospheres used in this study were generated with the ATLAS9 
code (R. L. Kurucz 1993, private communication).
A number of spectral lines, covering as many elements as possible,
were analyzed; most lines are largely unblended, so equivalent widths alone
are suitable for an abundance analysis.  In some instances, however, lines
are partially blended and spectrum synthesis was used. 

\subsection{The Fe I--Fe II Analysis and Stellar Parameters}

The neutral and singly ionized iron lines are used to derive basic
physical quantities for HD19994.  The lines employed are unblended
and have accurately measured gf-values.  These same lines have been
used in recent work by Smith et al. (2000) and the sources of the
gf-values are described in more detail there.  It is worth noting,
however, that the accuracy of most of these oscillator strengths are
now at the few percent level (Lambert et al. 1996).  Wavelengths,
excitation potentials, gf-values, and equivalent widths for HD19994
are listed in Table 1, for 20 Fe I and 4 Fe II lines.     

With a set of model atmospheres, Fe abundances can be derived for
various T$_{\rm eff}$'s, gravities, and microturbulent velocities.
A simultaneous fit, demanding the same Fe abundance from low- and
high-excitation ($\chi$$\sim$ 2-5 eV)
and weak and strong ($\sim$30-120 m\AA) Fe I lines, as
well as the Fe II lines, yields spectroscopic values of T$_{\rm eff}$,
log g, $\xi$, and Fe abundance.  Figure 1 illustrates part of
this process.
The slope of Fe-abundance (in units of log $\epsilon$(Fe)= log[N(Fe)/N(H)]
+ 12.) versus reduced equivalent-width (log W/$\lambda$) is plotted on
the x-axis, while the slope of log $\epsilon$(Fe) versus excitation
potential ($\chi$) is plotted on the y-axis.  The slopes are derived
from a linear least-squares fit to pairs of points of log $\epsilon$(Fe)
versus either log(W/$\lambda$) or $\chi$.  Various model atmosphere
effective temperatures and microturbulent velocities are shown in
Figure 1, with the
goal being to achieve no trend of Fe abundance with either $\chi$ 
or log(W/$\lambda$),
i.e. zero slopes in both axis.  The results illustrated are for the 
sample of Fe I lines and it is clear, that for this gravity (log g= 3.95),
T$_{\rm eff}$= 6030K and $\xi$=1.55 km s$^{-1}$ are the best solution, as
indicated by the large filled square at point 0,0.  The errorbars shown on
this point are derived from the linear least-squares fits to the relations
of Fe versus reduced equivalent-width and excitation potential.

In practice, this procedure is done for a variety of log g values, and
at only one particular log g will the Fe I and Fe II lines yield the same
Fe abundances.  For HD19994, the value of log g illustrated in Figure 1
(log g= 3.95) provides the same Fe I and Fe II abundances.  The 
spectroscopically derived best parameters for HD19994 are thus
T$_{\rm eff}$= 6030K, log g= 3.95, and $\xi$=1.55 km s$^{-1}$.  With
such high-S/N spectra as studied here, and  a sample of lines with
accurate gf-values, an internally consistent set of parameters can be
defined with uncertainties of about $\pm$20K in T$_{\rm eff}$, $\pm$0.05 in
log g, and $\pm$0.05 km s$^{-1}$ in $\xi$.  Of course, there will be 
systematic effects from both the choice of model atmosphere family and 
(probably) analysis code, however, these effects are now at about the
0.1 dex level for solar-type stars (with near-solar metallicities), as
can be seen by comparisons between various groups studying the same star.

In particular, HD19994 has been analyzed previously and comparisons can be made between 
the stellar parameters derived from different investigations.  From the large 
abundance study by Edvardsson et al. (1993), this star was found to have
T$_{\rm eff}$= 6104K, log g= 4.10, and $\xi$= 1.85 km s$^{-1}$.
Carretta, Gratton, \& Sneden (2000) have also included HD19994 in an
abundance analysis of solar-type stars and find T$_{\rm eff}$= 6015K,
log g= 3.94, and $\xi$= 1.47 km s$^{-1}$.  The agreement between the
various physical quantities derived for HD19994 from three different
studies is very good ($\sigma$(T$_{\rm eff})$= 47K, $\sigma$(log g)=
0.09 dex, and $\sigma$($\xi$)= 0.2 km s$^{-1}$). 

Furthermore, HD19994 is near enough to have a well-defined Hipparcos parallax
of $\pi$= 44.79$\pm$0.75 milli-arcseconds, thus, its distance is also
known accurately.  Given its distance and measured V-magnitude, an
absolute visual magnitude can be computed, and, with T$_{\rm eff}$=
6030K, the bolometric correction is small so an accurate luminosity
can be derived.  Given a luminosity, along with the stellar parameters,
a stellar mass can be derived from the relation
\begin{displaymath}
\frac{M_{*}}{M_{\odot}} = (\frac{L_{*}}{L_{\odot}})
(\frac{g_{*}}{g_{\odot}}) (\frac{T_{\odot}}{T_{*}})^{4},
\end{displaymath}
where M, L, g, and T are, respectively, the masses, luminosities,
surface gravities, and effective temperatures of the star (*) and
Sun ($\odot$).  Table 2 summarizes the physical properties for HD19994:
note that we derive a `spectroscopic mass' of M$_{*}$= 1.24M$_{\odot}$.
Use of the Carretta et al. (2000) stellar parameters (but same L$_{*}$)
would yield a near-identical mass of M$_{*}$=1.23M$_{\odot}$, while
the recent study by Allende-Prieto and Lambert (1999), using Hipparcos
parallaxes plus stellar model tracks, finds a mass of 1.28M$_{\odot}$ for
HD19994: all three of these studies find excellent agreement for the
stellar mass.   The earlier analysis by Edvardsson et al. (1993), in
which HD19994 had a slightly higher T$_{\rm eff}$ and log g, would
yield a somewhat higher mass of 1.68M$_{\odot}$, which is a bit on the
high side for a star classified as an F8V (although the use of the
Edvardsson et al. parameters would have only a minor effect on an abundance
analysis).  Santos, Israelian, \& Mayor (2001) have also analzed HD19994
and their stellar parameters are T$_{\rm eff}$= 6160K and log g =4.25:
these are both somewhat larger values than those derived by the studies
mentioned above.  Use of these particular values for effective temperature
and gravity, along with the luminosity derived from the Hipparcos parallax
(the relatively small differences in T$_{\rm eff}$ from the above studies
will lead to negligible differences in any bolometric corrections), would
yield a mass of 2.3M$_{\odot}$ for HD19994: this is much too large for  
its spectral type (F8V).  Both a larger T$_{\rm eff}$ and log g will lead
to, typically, larger derived abundances, so it is no surprise that
Santos et al. (2001) measure somewhat larger abundances ([Fe/H]= +0.23). 
Most of the more recent studies described above, however, indicate that the
parameters for HD19994 are secure and that an accurate abundance analysis
can be conducted on this star. 

With a relatively well-defined T$_{\rm eff}$ and luminosity, the mass of
HD19994 can also be estimated from stellar evolutionary model tracks.
Using the stellar models from Schaerer et al. (1993), with Z=0.02 (as
HD19994 is only modestly enhanced in metals with respect to the Sun)
we derive a mass for
HD19994 of 1.3M$_{\odot}$ and an age of 4.3Gyr, quite close to the age of
the Sun and solar system.

\subsection{Abundances Other Than Fe}

There are 21 elements, other than Fe, which have lines analyzed in this
spectrum of HD19994 (including boron which was analyzed by Cunha et al.
2000 using HST UV-spectra).  Table 3 contains the line identifications,
wavelengths, excitation potentials, gf-values, and measured equivalent 
widths in HD19994.
The sources of the
gf-values are taken either from Smith et al. (2000), who employed a
very similar set of lines in
an analysis of globular cluster giants using red spectra, or from
Gonzalez et al. (2001), who provide a large
compendium of abundances derived for stars with extrasolar planets. 
Both studies provide details about the gf-values, with most of the
Smith et al. oscillator strengths being measured values taken from the
literature (with a few solar values derived from a solar model generated
with the MARCS code), while the Gonzalez et al. gf-values are solar
ones, derived from an analysis of a reflected solar spectrum 
(Vesta) using a Kurucz ATLAS9 solar model.  Differences between the
solar gf-values derived from Smith et al. and Gonzalez et al. solar
analyses are small.

Some elements required spectrum synthesis, thus to both illustrate this
method of deriving abundances, as well as to demonstrate the overall quality
of the spectra, Figure 2 shows two regions which were synthesized in HD19994.
The [O I] line near 6300\AA\ and the Li I line near 6707\AA\ are shown,
with the filled circles being the observed data points and the solid
curves the synthetic spectra.  Note the telluric O$_{2}$ line near the
6300\AA\ [O I] line: rapidly-rotating hot-stars are observed to provide
templates to map the telluric lines and can be used to ratio out the O$_{2}$
lines, however, if the telluric lines do not interfere with fitting the
[O I] feature, no ratioing is done, as is the case here. 

In this study a direct abundance comparison
is made between the Sun and HD19994.  Abundances
are derived from the same sets of lines as those used for HD19994 and
the final elemental abundances found for HD19994 are then ratioed to
their respective solar values and abundances are then presented as
[X/H]. 
Final abundances are listed in Table 4.
The solar equivalent-widths (or spectra for spectrum synthesis) were measured 
from the solar flux atlas of Kurucz et al. (1984) and the lines analyzed
using an ATLAS9 model with T$_{\rm eff}$= 5777K, log g= 4.43, and
$\xi$=1.0 km s$^{-1}$ (as derived from this set of Fe I lines).  The
differences between the abundances derived here for the Sun and the
accepted photospheric values (e.g. Grevesse et al. 1996) are small. (See
discussion in Smith et al. 2000).  

Values of the abundances in HD19994, relative to the Sun, as [X/H] are
plotted in Figure 3.  The 22 elements studied here range in mass from
Li, B, or C, up to Eu and span a range in nucleosynthetic origins.
Other than a general overabundance of +0.13 dex in HD19994 (as indicated 
by the solid horizontal line, with the solar ratio of [X/H]=0.0 shown by the
horizontal dashed line), no other trends are apparent in this figure.  This
star does continue the tendancy, however, of planet-bearing stars being 
metal-rich.  Although not ``super-metal rich'', which is defined to be
at least +0.2 dex enriched, relative to the Sun, HD19994 is more metal-rich
than the Sun and is near the peak of the metallicity distribution found for
the stars with planets.  The abundance of [Li/H] is not plotted in Figure
3 as this element is very sensitive to stellar mixing processes and the
abundance of log $\epsilon$(Li)= 1.81 is not unusual for a late-F dwarf.

\bigskip
\section{Discussion}

Two aspects of the abundances in HD19994 will be discussed here:
are there overall abundance signatures in the
stars with large planets compared to stars not known to have large
planets, and are there any indications of chemical fractionation
occurring in possible accretion processes in stars with large planets. 

\subsection{Abundance Ratios in HD19994 Compared to Stars with Large Planets 
and Field Stars}

As pointed out in the Introduction, most abundance studies of stars with
large planets have found that these stars tend to have rather high 
metallicities: the most recent papers and summaries of work are by 
Gonzalez et al. (2001) and Santos et al. (2001).  
In the top panel of Figure 4 are shown the Fe-abundance distributions
for a volume-limited sample of solar-type stars, along with the same
distribution for the known stars
with planets that have been spectroscopically analyzed for their Fe
abundances.  The volume-limited sample is from 
Favata et al. (1997), while the planet-harboring stars are taken from
Gonzalez et al. (2001) and Santos et al. (2001).
These two studies include results for 30 stars, but we
show in Figure 4 only those stars for which oxygen abundances have also
been determined (26 stars).  The arrow shows the [Fe/H] 
abundance for HD19994, which is near the peak of the distribution for
planet-bearing stars.  To date, the parent stars with planets 
exhibit a distribution of Fe abundances skewed strongly towards larger
[Fe/H] values.  Since these two separate stellar samples have not been
chosen in the same way, there is the possibility of selection effects that
are not understood playing a role in the [Fe/H] distributions for the stars
with planets.
Nonetheless, the relatively large metallicities found for the planet-hosting
stars are interesting and the detailed abundance distributions in these
stars warrant a closer look.

The distribution of [O/H] abundances for planet-hosting stars is shown in
the bottom panel of Figure 4 (no such volume-limited distribution exists
for field solar-type stars).  Most of these oxygen abundances are corrected
values taken  
from the Gonzalez et al. (2001) study, and were obtained
from the IR O I triplet lines at 7774\AA.  These particular O I lines are
quite strong and may suffer from non-LTE or granulation effects.   
Edvardsson et al. (1993), in a large study of F and G stars, compared
LTE oxygen abundances derived from both the [O I] 6300\AA\ line (which forms
in LTE) and
the O I 7774\AA\ lines.  They determined empirical corrections that can
be applied to the 7774\AA\ O I abundances to bring them into agreement
with the [O I] 6300\AA\ line; this correction
is valid for the stellar parameters spanned by the stars with planets
studied by Gonzalez et al.  Thus, the O abundances shown in the bottom
panel of Figure 4 are derived from the LTE 7774\AA\ abundances, with
the Edvardsson et al. correction applied to them. 
The [O/H] abundance distribution is shifted slightly towards smaller values
than those for [Fe/H] by about 0.1 dex, with a somewhat larger fraction
of stars with [O/H]$\le$ 0.0.  Oxygen is shown as a comparison to iron
because, as discussed in the Introduction, the abundance of O may play an
important role in large-planet formation.

Comparisons of strategic abundance ratios are shown in Figures 5, 6, and
7 for various samples of field solar-type stars, plus HD19994, and other
stars with planets.  Both Gonzalez et al. (2001) and Santos et al. (2001) have 
previously conducted such
a comparison, however, here we select slightly different field-star samples as
comparisons, as well as add HD19994, and use oxygen as the fiducial metallicity
indicator instead of iron.  Oxygen abundances are somewhat more difficult
to measure than iron, with fewer O I lines than Fe I and Fe II, and with 
the stronger O I
lines suffering from possbile non-LTE effects (such as the O I 7774\AA\
triplet).  Although free from non-LTE effects, the [O I] 6300\AA\ is
quite weak and requires spectra of rather high-S/N.  Oxygen is, however,
a superior monitor of metallicity in stars with planets as it is not only
the most abundant element after H and He, but also plays a major role
in the formation of large planets, as these planets form initially around
icy cores, where H$_{2}$O and CO$_{2}$ dominate.  
Figure 5 compares the [O/Fe] ratios, versus [O/H], for known stars
with planets and a series of field-star samples of F, G, and K dwarfs.  
As discussed above, the O abundances for the Gonzalez et al. stars
are LTE abundances derived from the 7774\AA\ O I lines, with the
empirical corrections derived by Edvardsson et al. (1993) applied to them.
The other studies shown in Figure 5 either rely directly on the [O I]
6300\AA\ line (which is a first class abundance indicator [Lambert 1978]),
or use the O I 7774\AA\ lines with non-LTE or empirical corrections
,e.g. Edvardsson et al. (1993), applied to the LTE abundances.
No large differences exist between the various investigators
in the adoption of stellar parameters for near-solar metallicity
solar-type stars, thus all of the abundances shown in Figure 5 should be
on nearly the same scale.  This is born out by the large degree of
overlap between the different studies, and no apparent significant offsets.

There is nothing remarkable in Figure 5, except to note how well the
stars with planets fit into the overall trends of the field stars.  Above
a metallicity of $\sim$-0.15 in [O/H], both the field stars and stars with
planets show a rather flat run of [O/H], with small scatter for most
stars ($\sim$$\pm$0.1 dex).  Below this value of [O/H], the well-known 
increase in [O/Fe] with decreasing metallicity is apparent, again, with
most stars showing only small scatter from a mean trend.   Note that in
the distribution of [O/Fe] versus [O/H] in Figure 5, there is a fairly
well-defined lower envelope, with only a few outliers, however, there are 
conspicuous numbers of stars falling above the mean trend (i.e. large
values of [O/Fe] for a given [Fe/H]).  Feltzing \& Gustafsson's (1998) 
sample contains the largest fraction of these stars (and they specifically
concentrated on metal-rich stars), although all studies show some stars
in this region of the diagram.  By-and-large, however, the stars with
large planets do not stand out from the field distribution, except tending
to fall towards the high-metallicity end (as already discussed).  HD19994
falls  near the middle of the distribution of [O/Fe] and [O/H]
abundances, and appears normal in this regard. 

Another crucial element in the formation of ices in protoplanetary
disks is carbon, and comparisons using this element as a basis are
shown in Figure 6.  The top panel is [C/Fe] versus [O/H] while the
bottom panel is the ratio [C/O] versus [O/H].  Field-star samples with
both C and O are more limited than many other elements and the comparison
sample here consists of C abundances from Tomkin et al. (1995) combined
with O abundances in those same stars from Edvardsson et al. (1993).
Tomkin et al. adopted the same stellar parameters as Edvardsson et al.,
so these two studies should be self-consistent.  In addition, Tomkin et al.
used the C I lines near 7110\AA\, with many of the C abundances for the
stars with planets being derived from these same lines (as well as in 
HD19994), thus there do not appear to be any offsets between the field
star sample of stars not known to have planets and those stars known to
have planets.  
Although there are fewer field comparison stars, there again appears
to be nothing dramatic in a comparison of carbon in stars known to have
large planets, and those not known; HD19994 falls near the upper envelope
in both plots of the stars with planets, indicative of a very slightly 
larger C abundance, but not significantly so.

Finally, in Figure 7, the elements Na and Al are included in a comparison.
Of all the elements analyzed by Gonzalez et al. (2001), they found a
hint that Na might be slightly lower in the stars with large planets.
The top panel of Figure 8 shows [O/Na] versus [O/H] and there is a small
subset of the stars with planets that falls above the trend of [O/Na]
versus [O/H] as defined by the Edvardsson et al. (1993), Feltzing \&
Gustafsson (1998)  and Carretta et al. (2000) field-star results.
All of these various studies use some combination of the same Na I lines near  
5685\AA\ or 6155\AA\ and the different samples of abundances all
overlap to a large degree: there are no significant offsets from one study
to the next.

The modest enhancement of [O/Na] in some of the stars with planets 
is a combination
of slightly larger oxygen and slightly lower Na abundances.  Remarkably,
many of the Feltzing \& Gustaffson stars (which were chosen to be of high
metallicity) fall within this small 
grouping.  HD19994 falls near the low end of [O/Na] for the stars with
planets, but near the middle of the main trend in the field star samples.
We thus confirm the tentative conclusion from Gonzalez et al. that there
are a sizable fraction of stars with large planets that have slightly
lower Na abundances (and this result is strengthened by the fact that they
used Fe as the fiducial comparison element while we use O).  
A plot of [Na/Al], however,
in the bottom panel of Figure 7 shows that the stars with planets fall
right within the scatter of field stars from Edvardsson et al. (1993) and
Feltzing \& Gustafsson (1998). 

\subsection{Is There Evidence of Chemical Fractionation?}

Besides comparing overall abundance trends in stars with large planets
versus samples of stars not known to have planets, internal abundance
distributions can be a useful, and potentially more sensitive, diagnostic.  
As mentioned in the
Introduction, one possible explanation for the large metallicities
found in stars with large planets is that the parent stars accreted
some material, depleted in H and He, from the disk out of which the
planets formed.  Gonzalez et al. (2001) and Santos et al. (2001) reach
different conclusions concerning evidence for this accretion scenario.
Santos et al. find no trends of [Fe/H] versus stellar convective
envelope mass and conclude that accretion of fractionated material is
not enough to significantly affect the metallicities.  Gonzalez et al.,
on the other hand, suggest that accretion may play a role in the
observed [Fe/H] distribution.  A recent study by Laws \& Gonzalez (2001)
of 16 Cyg A and B finds small, but significant, differences in [Fe/H]
for the two members of this binary system, as well as a large difference
in their respective Li abundances.  They suggest that these abundance
differences could be due to the accretion of planetary material by
16 Cyg A, resulting in its somewhat larger [Fe/H] and [Li/H] values
relative to 16 Cyb B. 

An investigation into possible accretion can be probed further by searching
for abundance patterns that depend upon elemental condensation temperature,
T$_{\rm c}$.  If solid material is accreted by a star, in quantity, the
accretion might possibly occur in a rather high temperature environment
(i.e. close to the star).  In such a situation, refractory elements might
be added preferentially when compared to volatile elements.  
This idea was explored initially by Gonzalez (1997) in an abundance
analysis of $\upsilon$ And and $\tau$ Boo.  Gonzalez also searched for
possible evidence of a fractionation pattern in the Sun by comparing
differences between solar photospheric and meteoritic abundances as a
function of elemental condensation temperature, but did not detect a
significant signature.
In Figure 8 the abundances in HD19994
are plotted versus the elemental condensation temperature, with T$_{\rm c}$
taken from Lodder \& Fegley's (1998) table of T$_{\rm c}$'s for a solar
composition at a pressure of 10$^{-3}$ bar.  Changing the conditions
will change the absolute values of T$_{\rm c}$, but not the relative
order of the elements as shown in Figure 8.  If significant accretion
of material depleted in H and He occurred, there might also be other
fractionation patterns in the accreted
matter.  In such a case, trends might be detectable in the abundances
as a function of T$_{\rm c}$; in the case of HD19994 (Figure 8), no such
trend is apparent.  A trend in [X/H] versus T$_{\rm c}$ can
be quantified by a single number if a straight line is fit to the abundance
versus condensation temperature points.  The appearance, or lack,
of a slope in such a plot does not involve any physical arguments about
whether accretion of fractionated material should necessarily produce a
linear trend, but is simply a way of quantifying the abundance distribution
with T$_{\rm c}$ and comparing different stars.  A linear least-squares
fit to such a plot for HD19994 finds an insignificant negative slope of  
-6.1($\pm$9.8)x10$^{-6}$ dex/K, indicating no measurable trend 
whatsoever.  If material was accreted by HD19994 it was only depleted
measurably in H (and probably He). 

In this paper, the same type of analysis, looking for fractionation patterns, 
was also carried out for the abundances presented
by Gonzalez et al. (2001), where typically  15 elements were analyzed,
including such elements as C and O, which have low T$_{\rm c}$, S and Na
which have intermediate condensation temperatures, and
Fe, Ti, or Al, which have fairly high values of T$_{\rm c}$.  The derived
slopes are illustrated graphically in Figure 9, along with the result for
HD19994 obtained here.  The abundance work by Santos et al. (2001) does
not contain such a large number of elements, so his distributions are
not used in this sample comparison; the more elements sampled, and the
wider the range in T$_{\rm c}$, lead to more constrained slopes in [X/H]
versus T$_{\rm c}$.  The average of the Gonzalez et al.
slopes is shown by the vertical dashed line, with plus and minus one standard
deviation lines also shown, and there is a small positive 
offset from slope zero.
Five stars stand out in Figure 9, with the largest
positive slopes of [X/H] with T$_{\rm c}$: these stars
are HD52265, HD75289, HD89744, HD209458, and HD217107.  The slopes are
such that low condensation species, such as C, N, or O, are $\sim$ 0.2 dex
lower in abundance than the more refractory species, such as Ti, Sc, or
Al.  These stars are prime candidates for possibly having accreted
substantially fractionated material and additional, detailed abundance
analyses of these stars are encouraged.   The star with the most
negative slope in Figure 9 is HD37124, which also has the lowest iron
abundance with [Fe/H]= -0.41, while the star with the next lowest slope
is HD46375, also has a relatively low metallicity ([Fe/H]=-0.03)
for this sample of stars.  The lower slopes associated with lower values
of [Fe/H] suggests that there are chemical evolutionary effects, which
are not related at all with fractionation, and must be investigated.   

In order to probe the effects of chemical evolution on fitting slopes to
[X/H] versus T$_{\rm c}$, two samples of field stars, which covered a
wide range of elements, were used: Edvardsson et al. (1993) and
Feltzing \& Gustafsson (1998). 
Disk chemical evolutionary effects are illustrated in Figure 10,
where the parameterized slopes are plotted versus [Fe/H].  The stars from
Edvardsson et al. (1993) and Feltzing and Gustafsson (1998) that are 
used as ``control'' samples with which
to see effects from general chemical evolution are
plotted as open symbols in Figure 10 and identified as BDP (for Big Disk
Paper) and FG (Feltzing \& Gustaffson).  Stars from Gonzalez et al. (2001), 
as well as HD19994, are
shown as filled circles.  Error bars reflect the statistical uncertainty
in the derived slopes, with assumed abundance errors of 0.10 dex in
[X/H].  All three samples of stars use somewhat different mixtures of
elements, which could lead to small systematic differences in the derived
T$_{\rm c}$ slopes.  There is a large overlap, however, in the derived
slopes from all samples, and the three studies all include elements with
a broad range of condensation temperatures (especially O).  The large
degree of overlap suggests that any systematic effects between the studies
are smaller than the scatter. 

Concentrating on the BDP and FG points, there is a clear trend of a decreasing
slope in [X/H] versus T$_{\rm c}$ as [Fe/H] decreases: this is the
signature of Galactic chemical evolution in this diagram and not an
indication of chemical fractionation.  In these data, oxygen is one of
the most volatile species included, and carries a great
deal of weight in defining the slopes; as [O/Fe] rises towards lower
metallicities, it tends to produce a negative slope in a plot of [X/H]
versus T$_{\rm c}$.  A straight line
through the BDP points is shown to illustrate the general trend in the
field stars.  As the abundances used in the analyses are all relative to
the Sun, the fitted straight line goes nearly through the zero-zero
point as it should.  In general, the stars with large planets show a
remarkable overlap with the general field samples: most of the stars with 
planets fall within the scatter of the BDP and FG results, suggesting no
measurable differences in the overall abundance distributions with condensation
temperature at a given [Fe/H].  Of course, not all of the Edvardsson et
al. and Feltzing and Gustafsson stars have been searched for large 
planets, and there remains the possibility that fractionation effects
lie in small numbers of stars in both samples.  To search for
selective accretion of refractory species, the target stars should have 
large positive slopes, and there is such a 
small subset of 5-6 stars in Figure 10 that cluster at larger [Fe/H] 
and fall above the general scatter: these include the same 5
stars noted from Figure 9.  These stars show an abundance pattern not
seen in the other stars due to general, overall chemical evolution of
the Galaxy.  
A total of 6 stars stand out from combining results from both 
Figure 9 and 10 as having slopes well above the mean
trend: these stars are HD52265, HD75289, HD89744, HD120136,
HD209458, and HD217107.  Abundances as a function of T$_{\rm c}$ are
shown for two stars from this group (HD75289 and HD209458)
in Figure 11.  As discussed previously, the straight-line fits to the
abundances are not met to imply that there is a real linear relation
between [X/H] and T$_{\rm c}$, but are used to characterize the
abundance distribution by a single number (the slope).  Both stars
exhibit a pattern in which the more volatile elements (C, N, O, S, and Zn)
are $\sim$ 0.3 dex lower in abundance than the more refractory
elements (Si, Ti, Ca, and Sc). 
This is clearly not the definitive study,  as a detailed, uniform
analysis must be applied to samples of stars with planets and stars
known not to have large planets, yet the stars noted above 
should be subjected to very careful abundance
analyses, as they exhibit the clearest signatures of potential accretion
of fractionated material.  Note that two of the Feltzing \& Gustafsson (1998)
stars also fall within this selective group (this is about 10\% of the
FG stars plotted here and we only used those FG stars in which they determined
O abundances): HD110010 and HD137510 (these stars should be prime candidates
for having closely orbiting planets).  This percentage of stars in the
FG sample that might have planets is in approximate agreement with the
recent study by Laughlin (2000), who argues that $\sim$10\% of solar-type
field stars with [Fe/H]$\ge$ +0.2 might have large planets. 

Other than larger than normal slopes in plots of [X/H] versus T$_{\rm c}$,
do the abovementioned stars exhibit any other distinctive properties?
Physical parameters of the detected planets in extra-solar systems are 
shown in three panels
of Figure 12, with the stars with the larger T$_{\rm c}$ slopes plotted
as filled circles.  In the top two panels of Figure 12, the companion
masses (Msin($\iota$)) are plotted versus eccentricity ({\sl e}) and semi-major
axis ({\sl a}): the stars with possible accretion signatures stand out as having
smaller orbital separations, as well as possibly somewhat smaller
eccentricities and companion masses (although the masses are only 
lower limits, as the inclinations are not, in general, known).  In the
bottom panel the eccentricities and semi-major axes are compared, and
the stars showing accretion evidence are well segregated to the small {\sl e},
small {\sl a} part of this plot.  

A striking difference between the stars with the largest
T$_{\rm c}$-slopes and the others can be seen in the respective
distributions of their companion's semi-major axes.  
Figure 13 shows cumulative fractional
distributions of the values of {\sl a} (with the cumulative fraction being
the fraction of stars in that sample having companions which have
{\sl a} less than a particular
value).  In the top panel, the entire sample of 30 stars is shown, while
in the bottom panel, the sample is divided into the 6 stars falling furthest
from the general trend of T$_{\rm c}$-slope versus [Fe/H] (filled circles),
and the other 24 (open circles).  The stars showing the most evidence of
possible fractionated accretion are biased strongly towards having
close companions and the two distributions are clearly different.
Could these two differing distributions be caused by random sampling?
As a check on this, random 6-point samplings of the entire 30-point
data set were conducted and subsequent cumulative fractional distributions
were computed for each random set.  Doing this exercise 5000 times it
was found that a cumulative fractional distribution rising as rapidly
as that found for the 6 stars with the large T$_{\rm c}$-slopes occurs
only 5\% of the time.  It is unlikely that the observed distribution has
arisen by chance; there is a link between the abundance
distributions and small separations of large planets.

The ``dynamical'' properties of the orbits are completely
independent from abundance analyses and suggest that there must
be a physical process at work here.  
Lower eccentricities may be a fossil
signature of a protoplanetary disk, as small eccentricities are a
requirement of the standard model for giant-planet formation, resulting
from the gradual accretion of small solid particles in a disk, followed
by gravitational accretion of gas.   Smaller orbital
separations suggest the possibility of more interaction between
planet, disk, and star, as the planet presumably migrated inwards to
its current position around its parent star when it was forming.  
It is worth noting that among the stars which exhibit signatures of
fractionated accretion is HD209458.  This star has the only companion
whose mass has been well-determined by transit observations: M$_{\rm comp}$=
0.63M$_{\rm J}$ (Charbonneau et al. 2000).  This mass definitely
classifies it as a planet, well below the deuterium-burning limit.

Another property of the stars that can be examined is their mass.
Again, the six star-planet systems with the largest slopes of [X/H]
versus T$_{\rm c}$ are somewhat distinctive relative to the other
star-planet systems.  The 24 systems that do not exhibit T$_{\rm c}$
slopes noticably different from field-star samples of stars not known
to have planets have a mean stellar mass of 1.06M$_{\odot}$ (with a
standard deviation of 0.12M$_{\odot}$), while the 6 stars exhibiting
the largest T$_{\rm c}$ slopes have a mean mass of 1.23M$_{\odot}$
(with a standard deviation of 0.17M$_{\odot}$).  These mass estimates
are taken from the papers by Gonzalez et al. (2001) and Santos
et al. (2001) and have been estimated using stellar evolutionary tracks. 
The suggestive link here would be that the more massive stars have
smaller mass convective envelopes and would thus be easier to pollute
with accreted material.  It is worth noting that, for an age of
10$^{8}$ years and models from D'Antona \& Mazzitelli (1994), the
convective envelope mass goes to near-zero above a mass of 
$\sim$1.15M$_{\odot}$ and 5 of the 6 stars with significant
T$_{\rm c}$ slopes lie above this limit.  The one exception is
HD217107, which is estimated to have a mass of 0.98M$_{odot}$, however,
there are at least 0.1M$_{\odot}$ uncertainties in all of these estimates.
There are two systems that do have high masses and closely orbiting 
planets, but do not show any trend of [X/H] with condensation
temperature: HD9826 and HD38529.  Both of these systems have their
own unique properties within this sample.  HD9826, with a mass of
1.3M$_{\odot}$, has three large planets with semi-major axes ranging
from 0.05 to 2.5AU, with the two outer planets in eccentric orbits; the
dynamical history of this system could be quite complex.
With a mass of 1.5M$_{\odot}$, HD38529 has one of the largest masses in
the sample and has a planet at a distance of 0.13AU, however, it is also
one of the most highly evolved stars in the sample.  It lies well above
the main-sequence, by $\sim$1 magnitude, so if it had accreted any material
early in its history, a combination of a deepening convective envelope
and possible mass loss could have erased a thin veneer of accreted
matter. 

Because the data used to obtain the slopes illustrated in Figure 9
are quite heterogeneous, definitive conclusions must await further
observations; the prime motivation here is to draw attention
to these stars.  Within this sample, those stars with the largest
positive slopes of [X/H] versus T$_{\rm c}$ tend to have giant planets in 
near-circular orbits well within 1AU.  Further detailed abundance analyses
of these stars are in order, as well as results for more stars which have been 
observed and found not to have large planets in close orbits as a
comparison sample.  Elements
having low- to intermediate-condensation temperatures, such as S, Cu, or Zn 
should be included in the analyses, as well as the abundant trio C, N, and
O.

\bigskip
\section{Conclusions}

An abundance analysis covering some 22 elements has been presented for
HD19994, another star now known to harbor a large planet.  As found
in previous studies, e.g. Gonzalez et al. (2001) or Santos et al. (2001),
where the stars with large planets tend to be metal-rich, HD19994 is
slightly metal-rich, with [Fe/H]= +0.09$\pm$0.05.  An average of all 22
elements finds [X/H]=+0.13.  No obvious trends of nuclear origin are
found in the abundances of HD19994, nor is there a trend with elemental
condensation temperature, which might occur if the star had accreted
substantial amounts of chemically fractionated material from a
protoplanetary disk.  A comparison of the abundance distributions versus 
T$_{\rm c}$ in the sample of stars with planets from Gonzalez et al. (2001)
finds 6 stars which are the most likely candidates for having possibly 
accreted fractionated material.  It is also found that these 6 candidate
stars share the dynamical system property of having quite small orbital 
separations.  In addition, these 6 stars tend to be the more massive stars
known to have planets. 
Continuing abundance analyses of a
broad range of elements in all stars known to have planets, compared to
a sample of stars known not to have large planets, will shed light on
whether the formation of large planets can measurably alter the apparent
metallicities of their parent stars.  In addition, such work will 
help determine whether the formation of large planets is more likely to 
occur in a metal-rich environment. 

We thank the staff of McDonald Observatory.  The referee, Guillermo
Gonzalez, is to be thanked for providing a careful reading of the
initial paper and for suggesting some useful comments which improved the
final version.
This research is supported in part 
by the National Science Foundation (AST99-87374).

\clearpage

\figcaption[Figure1.ps]{An illustration of the derivation of the fundamental
stellar parameters T$_{\rm eff}$, log g, and microturbulence ($\xi$).
The slope of log $\epsilon$(Fe I) versus excitation potential ($\chi$) is
shown as a function of the slope of log $\epsilon$(Fe I) versus
log(W/$\lambda$); zero slopes in both axes indicates the best values
of T$_{\rm eff}$ and $\xi$ (identified by the filled square).  Differing
model effective temperatures and micorturbulent velocities are shown by
the connected curves, with all of this illustrated for a single model
surface gravity (log g = 3.95).  Singly ionized iron, Fe II, is most
sensitive to log g, and the procedure shown is carried out for a range
of values of log g.  The example shown here of log g= 3.95 yields the
same Fe I and Fe II abundances for the best values of T$_{\rm eff}$ and
$\xi$. 
\label{fig1}}

\figcaption[Figure2.ps]{Sample observed and synthetic spectra for two
small regions in HD19994.  Two weak features used for deriving abundances
are shown, with [O I] 6300\AA\ in the top panel and Li I 6707\AA\ in
the bottom panel.  Three different O and Li abundances are shown for
each set of synthetic spectra in each panel, repsectively, as well as
syntheses with no O (top) and no Li (bottom).  Extra broadening (beyond
instrumental) was required to fit the line shapes in HD19994, with a
typical velocity width of 6 km s$^{-1}$ needed; at this spectral
resolution, it is not possible to decompose this broadening into
macroturbulence plus rotation.  Note the telluric O$_{2}$ line in the
6300\AA\ region, which was not removed as it had no effect on the [O I]
fits. 
\label{fig2}}

\figcaption[Figure3.ps]{Abundances in HD19994 as a function of element
number, with values plotted as [X/H] relative to solar.  A solar
abundance ratio is indicated by the horizontal dashed line ([X/H]= 0.0),
while the solid horizontal line is the average value in HD19994 for
all elements shown ([X/H]= +0.13).
\label{fig3}}

\figcaption[Figure4.ps]{Iron and oxygen abundance frequency distributions
for a sample of solar-type stars and stars with planets.  The top panel
shows [Fe/H] distributions for a volume-limited sample of stars from
Favata et al. (1997) and a sample of stars with planets from Gonzalez
et al. (2001), Santos (2001), and HD19994 from this study (the arrows
indicate the abundances for HD19994).  Only the stars with planets that
have also had oxygen abundances measured are plotted.  The Fe distribution
of the stars with planets is skewed strongly towards rather large values
of [Fe/H].  The bottom panel shows [O/H] for the same planet-harboring
stars from the top panel (no volume-limited sample of field stars with
O abundances exists).  The [O/H] distribution is slightly different than
[Fe/H], being shifted somewhat ($\sim$0.1 dex) to lower values. 
\label{fig4}}

\figcaption[Figure5.ps]{The behavior of [O/Fe] with [O/H] (``metallicity'')
in several solar-type stars, including stars with planets.  The general
trend of Galactic disk chemical evolution is visible, with [O/Fe] being
relatively flat from [O/H]$\sim$ -0.15 to +0.20, and increasing below
[O/H]= -0.20.  The distributions of the various field-star samples overlap
very well.  In general, the stars with planets do not noticeably segregate
from the general population of field stars.  
\label{fig5}}

\figcaption[Figure6.ps]{Comparisons of [C/Fe] and [C/O] as a function of 
[O/H].  The stars with planets do not stand out noticeably from the field
F and G stars with measured C and O abundances.
\label{fig6}}

\figcaption[Figure7.ps]{Na and Al are compared to O in stars with planets
and a number of other field-star studies.  As discerned by Gonzalez et al.
(2001), a small group of stars with planets stand out as having somewhat
larger [O/Na] ratios at high values of [O/H].  A substantial fraction of
the high-metallicity Feltzing and Gustafsson (1998) sample also fall in
this region.  In terms of [Na/Al] ratios, the stars with planets fall
right within the field-star trends with little scatter. 
\label{fig7}}

\figcaption[Figure8.ps]{Abundances, relative to solar, in HD19994 plotted
versus the elemental condensation temperature, T$_{\rm c}$.  Values for
T$_{\rm c}$ are taken from Lodders \& Fegley (1998), and the elemental
names are indicated.  No trend of [X/H] with T$_{\rm c}$ exists in
HD19994, indicating a near-uniform abundance of all elements ($\sim$$\pm$
0.1 dex) relative to the Sun.  The solid line shows a linear least-squares
fit to these data, with the result that there is no measurable trend of
abundance with T$_{\rm c}$.  
\label{fig8}}

\figcaption[Figure9.ps]{Slopes derived from [X/H] versus T$_{\rm c}$ for
stars with planets from Gonzalez et al. (2001) and for HD19994.  The average
value of the slopes, with $\pm$1$\sigma$ to this average, are shown by
the vertical dashed lines.  The error bars on the individual points are
calculated from the goodness-of-fit to a straight line and individual
errors in [X/H] of 0.1 dex.  The two stars with the lowest (most negative)
slopes are two of the most metal-poor stars in the sample.  A small group
of 5-6 stars falls at higher values for the slopes: these stars are
possible candidates for fractionated accretion. 
\label{fig9}}

\figcaption[Figure10.ps]{Derived slopes of [X/H] versus T$_{\rm c}$ are
plotted as a function of [Fe/H] for three samples of stars (Edvardsson
et al. 1993--BDP; Feltzing \& Gustafsson 1998--FG; stars with planets from
Gonzalez et al. 2001 and HD19994--SWP).  There is a trend of decreasing
T$_{\rm c}$-slope with lower [Fe/H] caused by general chemical evolution
in the disk (due mostly to the increase of [O/Fe] as [Fe/H] decreases).
A small group of SWP points exist with larger T$_{\rm c}$-slopes, as well
as two FG high-metallicity stars.  These stars are segregated from the
general trend and are suggested to be the best candidates to have undergone
possible fractionated accretion. 
\label{fig10}}

\figcaption[Figure11.ps]{Abundances on the scale [X/H] versus elemental
condensation temperature (T$_{\rm c}$) for two stars from the six that
were identified as having the largest trends of [X/H] versus T$_{\rm c}$.
In both of these stars, the abundances of the more volatile elements are 
$\sim$0.3 dex lower than those from the more refractory species.  This
difference in abundance is significant and suggests 
selective mass accretion of refractory elements, possibly
through the addition of solid material into their outer convective
envelopes.   
\label{fig11}}

\figcaption[Figure12.ps]{Orbital properties and masses
(Msin($\iota$)) of the planets around stars exhibiting the largest slopes of 
[X/H] with
T$_{\rm c}$ (filled circles), and those stars falling closer to the
general trend of T$_{\rm c}$-slope versus [Fe/H] (open circles).  The top
two panels show Msin($\iota$) versus semi-major axis ({\sl a}) and
eccentricity ({\sl e}), respectively.  The most apparent trend in these two
panels is that the stars with large T$_{\rm c}$-slopes tend to have nearby
companions.  The bottom panel isolates {\sl e} versus {\sl a} and the stars with
large T$_{\rm c}$-slopes are clearly confined to the left part of the
plot ({\sl a} $\le$ 1 AU).  
\label{fig12}}

\figcaption[Figure13.ps]{Cumulative fractional orbital separations,
for all star--planet systems with measured abundances, are shown in 
the top panel.  The cumulative fraction is the fraction of the sample that
has an orbital separation, {\sl a}, less than a particular value of {\sl a}.
In the bottom panel the sample is split into those stars with the larger
T$_{\rm c}$-slopes, and the other systems which follow the general field disk 
trend of T$_{\rm c}$-slope versus [Fe/H].  The distributions are
markedly different: stars with large slopes (and are considered to be the
best candidates to have undergone fractionated accretion) have companions
that are significantly closer, on average, than stars showing no strong
evidence of fractionated accretion.   
\label{fig13}}

\end{document}